\documentclass[showpacs,preprintnumbers,amsmath,amssymb]{revtex4}
\usepackage{stmaryrd}
\usepackage{txfonts}
\usepackage{amssymb}

\usepackage{graphicx}
\usepackage{dcolumn}
\usepackage{bm}

\begin{document}

\preprint{APS/123-QED}

\title{Lepton-Number Violating Decays of Heavy Mesons}

\author{Jin-Mei Zhang}
 \email{jinmeizhang@tom.com}
\affiliation{Department of Physics, Harbin Institute of
Technology, Harbin 150001, China. } \affiliation{ Xiamen Institute
of Standardization, Xiamen 361004, China. }
\author{Guo-Li Wang}
 \email{gl_wang@hit.edu.cn}
\affiliation{Department of Physics, Harbin Institute of Technology,
Harbin 150001, China. }

\date{\today}

\begin{abstract}
The experimental observation of lepton-number violating processes
would unambiguously indicate the Majorana nature of neutrinos.
Various $\Delta L$ = 2 processes for pseudoscalar meson $M_1$
decays to another pseudoscalar meson $M_2$ and two charged leptons
$\ell_1$, $\ell_2$ ($M_1^+\rightarrow \ell_1^+\ell_2^+M_2^-$) have
been studied extensively. Extending the existing literature on the
studies of these kinds of processes, we consider the rare decays
of heavy mesons to a vector meson or a pseudoscalar meson. These
processes have not been searched for experimentally, while they
may have sizable decay rates. We calculate their branching
fractions and propose to search for these decay modes in the
current and forthcoming experiments, in particular at the LHCb.
\end{abstract}

\pacs{13.25.Ft, 13.25.Hw, 14.40.Lb, 14.40.Nd, 14.60.St}

\maketitle

\section{\label{sec:level1}Introduction}

The neutrino oscillation experiments have proved that neutrinos are
massive \cite{Kam,SNO,NEMO,Barger}. However, the nature of neutrino
masses is still one of the main puzzles  in contemporary particle
physics, {\it i.e.}, are neutrinos Dirac or Majorana particles? As
we all known that the Majorana mass term violates lepton number by
two units ($\Delta L$ = 2). Thus, the unambiguous answer to the
question above is the experimental observation of a lepton-number
violating (LV) process.

Various $\Delta L$ = 2 processes have been studied in the literature
\cite{Ali,Tao,Atre,Claudio,Rodejohann}. Among them Atre {\it et al.}
\cite{Tao} have studied 36 LV processes from $K$, $D,\ D_s,$ and $B$
decays, generically written by:
\begin{equation}
M_1^+\rightarrow \ell_1^+\ell_2^+M_2^-,
\end{equation}
where $M_i^{\pm}$ and $\ell_i^+$ $(i=1, 2)$ denote charged
pseudoscalar mesons and leptons, respectively.

Most of these processes have been searched for and the
non-observation in the current experiments set the bounds on
branching fractions. In turn, they led to some stringent constrains
on the mixing parameters between Majorana neutrino and charged
lepton directly. However, there are still more LV heavy meson decay
modes that have not been studied experimentally, that may have
sizable branching fractions in theory. In particular, heavy mesons
(with $c,b$ flavors) are easier to identify and the LHCb experiments
will provide us with a large data sample. So as an extension of
current existing calculations, we explore some new $\Delta L=2$
decay modes in this paper. We mainly consider the rare decays of
heavy mesons $D, D_s,$ and $B$ to vector meson final states. Since
the LV heavy meson decay modes under our consideration have no
experimental results, we cannot extract the mixing parameters
through these decay channels like Atre {\it et al.} did.

However, those processes considered by Ali \cite{Ali} and Atre
\cite{Tao} are clearly correlated with the decay modes under our
consideration, with the same mixing parameters specified by the
charged lepton flavors. We thus adopt the numerical values of mixing
parameters extracted from Ref.~\cite{Tao}, and the decay widths and
branching fractions of heavy mesons for our processes can be
predicted correspondingly. We choose the strongest constrains on
mixing parameters from Ref.~\cite{Tao} as input in our study, in
order to be conservative.

we mainly consider the heavy pseudoscalar meson $D,\ D_s,$ and $B$
to vector meson final states. From theoretical point of view, the
decays of vector mesons may have different and uncorrelated rates
from that of the pseudoscalars if there is other type of new
physics, like a heavy particle exchange of either a
pseudoscalar/scalar or a vector boson. Thus, it is well motivated
to carry out the complementary searches for all of the availabel
final states.

The paper is organized as follows. In
Sec.~\uppercase\expandafter{\romannumeral2} we outline the useful
formulas to set the general notation. We list the constraints on
mixing parameters and give the Monte Carlo sampling of the branching
fractions as a function of the heavy neutrino mass in
Sec.~\uppercase\expandafter{\romannumeral3}, and draw our conclusion
in Sec.~\uppercase\expandafter{\romannumeral4}.

\section{The General Formalism For Lepton-Number Violating Decay}

\begin{figure}
\centering
\includegraphics[width=1.8in]{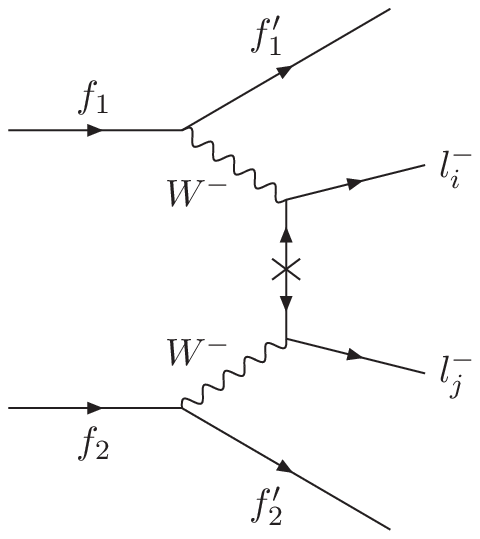}
\caption{\label{fig:feynman}Feynman diagram corresponding to the
$\Delta L = 2$ processes via Majorana neutrino exchange.}
\end{figure}

The simplest renormalizable extension of the standard model (SM) to
generate neutrino Majorana masses is to introduce $n$ right-handed
SM singlet neutrinos $N_{bR}\ (b = 1,2,\cdot\cdot\cdot,n)$.
Therefore, the complete neutrino mass sector is composed of both
Dirac masses that produced via the Yukawa couplings to the Higgs
doublet in the SM, and possible heavy Majorana mass term
$\frac{1}{2}\sum\limits_{b,b^{\prime
}=1}^{n}\overline{N_{bL}^c}B_{bb^{\prime}}N_{b^{\prime}R} + h.c.$ .

In terms of the mass eigenstates, the gauge interaction lagrangian
of the charged currents now has the following form:
\begin{eqnarray}
\mathcal
{L}=-\frac{g}{\sqrt{2}}W_\mu^+\left(\sum\limits_{\ell=e}^\tau
\sum\limits_{m=1}^3U_{\ell m}^{\ast }\overline{\nu_m}\gamma^\mu
P_L\ell+\sum\limits_{\ell=e}^\tau\sum\limits_{m^\prime=4}^{3+n}V_{\ell
m^{\prime}}^{\ast}\overline{N_{m^{\prime}}^c} \gamma^\mu
P_L\ell\right)+h.c.
\end{eqnarray}
where $P_L=\frac{1}{2}(1-\gamma_5)$, $\nu_m (m = 1,2,3)$ and
$N_{m^\prime} (m^\prime = 4,\cdots,3+n)$ are the mass eigenstates,
$U_{\ell m}$ is the mixing matrix between the light flavor and light
neutrinos, and $V_{\ell m^\prime}$ is the mixing matrix between the
light flavor and heavy neutrinos.

The basic process with $\Delta L = 2$ shown in
Fig.~\ref{fig:feynman} with exchange of  two virtual SM $W$ bosons
can be generically expressed by:
\begin{eqnarray}
W^-W^- \rightarrow \ell_1^-\ell_2^-\,,
\end{eqnarray}
where $\ell_{l,2}=e,\mu,\tau$. The process can occur only if
neutrinos are Majorana particles. Unfortunately, the transition rate
of this $\Delta L = 2$ process encounters a severe suppression
either due to the small neutrino mass like $m_{\nu m}^2/M_W^2$, or
due to the small mixing
$\left|V_{\ell_1m^{\prime}}V_{\ell_2m^{\prime}}\right|^2$. However,
when the heavy neutrino mass is kinematically accessible, the
process may undergo a resonant production of the heavy neutrino,
thus substantially enhancing the transition rate. The resonant
contributions of heavy Majorana neutrinos to $\Delta L = 2$
processes involving two charged leptons and another pseudoscalar
meson have been considered in the Ref.~\cite{Tao}. In this paper, we
extend the study of the heavy Majorana neutrinos to $\Delta L = 2$
processes involving two charged leptons and a vector meson final
state.

\begin{figure}
 \centering
\includegraphics[width=4.0in]{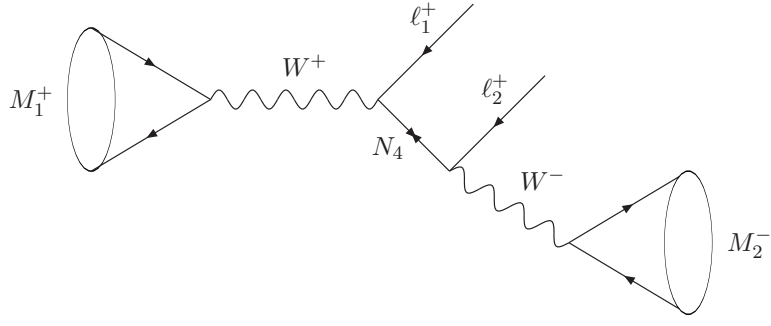}
\caption{\label{fig:decay}Feynman diagram corresponding to the
lepton-number violating decays $M_1^+(q_1) \rightarrow
\ell_1^+(p_1)\ell_2^+(p_2)M_2^-(q_2)$.}
\end{figure}

The Feynman diagram for the LV decay of heavy meson $M_1$ into two
charged leptons $\ell_1,\ \ell_2$ and another meson $M_2$:
\begin{eqnarray}\label{ntr1}
M_1^+(q_1)\rightarrow \ell_1^+(p_1)\ \ell_2^+(p_2)\ M_2^-(q_2)\,.
\end{eqnarray}
is shown in Fig.~\ref{fig:decay}. According to narrow-width
approximation, the tree level decay amplitude when $M_2$ is a
pseudoscalar meson is given as \cite{Tao}:
\begin{eqnarray}\label{ntr2}
i\mathcal
{M}^P=2G_F^2V_{M_1}^{CKM}V_{M_2}^{CKM}f_{M_1}f_{M_2}V_{\ell_14}
V_{\ell_24}m_4\left[\frac{\bar{u}_{\ell_1}\not\!{q_1}\not\!{q_2}
P_Rv_{\ell_2}}{(q_1-p_1)^2-m_4^2+i\Gamma_{N_4}m_4}\right]
+(p_1\leftrightarrow p_2)\,,
\end{eqnarray}
when $M_2$ is a vector meson, the tree level decay amplitude can be
written as \cite{Tao}:
\begin{eqnarray}\label{ntr3}
i\mathcal
{M}^V=2G_F^2V_{M_1}^{CKM}V_{M_2}^{CKM}f_{M_1}f_{M_2}V_{\ell_14}
V_{\ell_24}m_4m_{M_2}\left[\frac{\bar{u}_{\ell_1}
\not\!{q_1}\not\!{\epsilonup^\lambda(q_2)}P_Rv_{\ell_2}}
{(q_1-p_1)^2-m_4^2+i\Gamma_{N_4}m_4}\right]+(p_1\leftrightarrow
p_2)\,,
\end{eqnarray}
where $G_F$ is Fermi constant; $V_{M_i}^{CKM}$ is the
Cabibbo-Kobayashi-Maskawa (CKM) matrix elements; $f_{M_i}$ is the
decay constant for meson $M_i$; $q_1, q_2,\ p_1,\ p_2$ are the
momenta of mesons $M_1,\ M_2$ and leptons $\ell_1,\ \ell_2$,
respectively. Here we consider the case when only one heavy Majorana
neutrino is kinematically accessible and denote it by $N_4$, with
the corresponding mass $m_4$ and mixing with charged lepton flavors
$V_{\ell4}$.  $\Gamma_{N_4}$ is the total decay width of the heavy
Majorana neutrino, summing over all accessible final states. Then
the partial decay width $\Gamma_{\not\!{L}}^{M_1}$ and the
normalized branching fraction
$BR=\Gamma_{\not\!{L}}^{M_1}/\Gamma_{M_1}$ for the LV process
Eq.~(\ref{ntr1}) can be calculated by the decay amplitude. Following
the approach of Ref.~\cite{Tao}, we will take the mixing parameter
$V_{\ell 4}$ and the mass $m_{4}$ as phenomenological parameters.

\section{Monte Carlo Sampling For Lepton-Number Violating Decays}

\begin{table*}
\caption{\label{tab:table1}The decay modes of the leptonic-number
violating decays $M_1^+(q_1) \rightarrow
\ell_1^+(p_1)\ell_2^+(p_2)M_2^-(q_2)$ and the ranges of the mixing
parameters $|V_{\ell_14}V_{\ell_24}|$. The rages are calculated base
on the correlative decay modes from the Ref.~\cite{Tao}.}
\vspace{5mm}
\begin{tabular}{|c|c|c|c|}
\hline Decay mode&Mixing element $|V_{\ell_14}V_{\ell_24}|$&Range of
$m_4$(MeV)&Range of\ $|V_{\ell_14}V_{\ell_24}|$\\
\hline
$D^+\rightarrow e^+e^+\rho^-$& &\ \ 776 - 1869&\ \ 0.002 - 0.998\\
$\ \ D^+\rightarrow e^+e^+K^{*-}$& &\ \ 892 - 1869&\ \ 0.002 - 0.998\\
$D_s^+\rightarrow e^+e^+\rho^-$& &\ \ 776 - 1968&\ \ 0.023 - 0.992\\
$\ \ D_s^+\rightarrow e^+e^+K^{*-}$& &\ \ 892 - 1968&\ \ 0.022 - 0.992\\
$\ B^+\rightarrow e^+e^+D^-$&$|V_{e4}|^2$&1870 - 5278&\ \ 0.236 - 0.992\\
$\ B^+\rightarrow e^+e^+D_s^-$& &1969 - 5278&\ \ 0.231 - 0.992\\
$\ \ B^+\rightarrow e^+e^+D^{*-}$& &2011 - 5278&\ \ 0.231 - 0.992\\
$\ \ B^+\rightarrow e^+e^+D_s^{*-}$& &2113 - 5278&\ \ 0.232 - 0.992\\
\hline
$D_s^+\rightarrow \mu^+\mu^+\rho^-$& &\ \ 882 - 1863&\ \ 0.002 - 0.993\\
$\ B^+\rightarrow \mu^+\mu^+D^-$& &1975 - 5173&\ \ \ \ 0.28 - 0.971\\
$\ B^+\rightarrow \mu^+\mu^+D_s^-$&$|V_{\mu4}|^2$&2074 - 5173&\ \ \ \ 0.28 - 0.971\\
$\ \ B^+\rightarrow \mu^+\mu^+D^{*-}$& &2116 - 5173&\ \ \ \ 0.28 - 0.971\\
$\ \ B^+\rightarrow \mu^+\mu^+D_s^{*-}$& &2218 - 5173&\ \ 0.327 - 0.971\\
\hline
$D^+\rightarrow e^+\mu^+\rho^-$& &\ \ 776 - 1869&\ \ \ \ 0.02 - 0.497\\
$\ \ D^+\rightarrow e^+\mu^+K^{*-}$& &\ \ 892 - 1869&\ \ 0.023 - 0.497\\
$D_s^+\rightarrow e^+\mu^+\rho^-$& &\ \ 776 - 1868&\ \ \ \ 0.02 - 0.497\\
$\ \ D_s^+\rightarrow e^+\mu^+K^{*-}$& &\ \ 892 - 1968&\ \ 0.023 - 0.497\\
$\ B^+\rightarrow e^+\mu^+D^-$&$|V_{e4}V_{\mu4}|$&1870 - 5278&0.242 - 0.49\\
$\ B^+\rightarrow e^+\mu^+D_s^-$& &1969 - 5278&0.236 - 0.49\\
$\ \ B^+\rightarrow e^+\mu^+D^{*-}$& &2011 - 5278&0.236 - 0.49\\
$\ \ B^+\rightarrow e^+\mu^+D_s^{*-}$& &2113 - 5278&0.236 - 0.49\\
\hline
\end{tabular}
\end{table*}

\begin{figure}[tb]
 \centering
\includegraphics[width=5.0in]{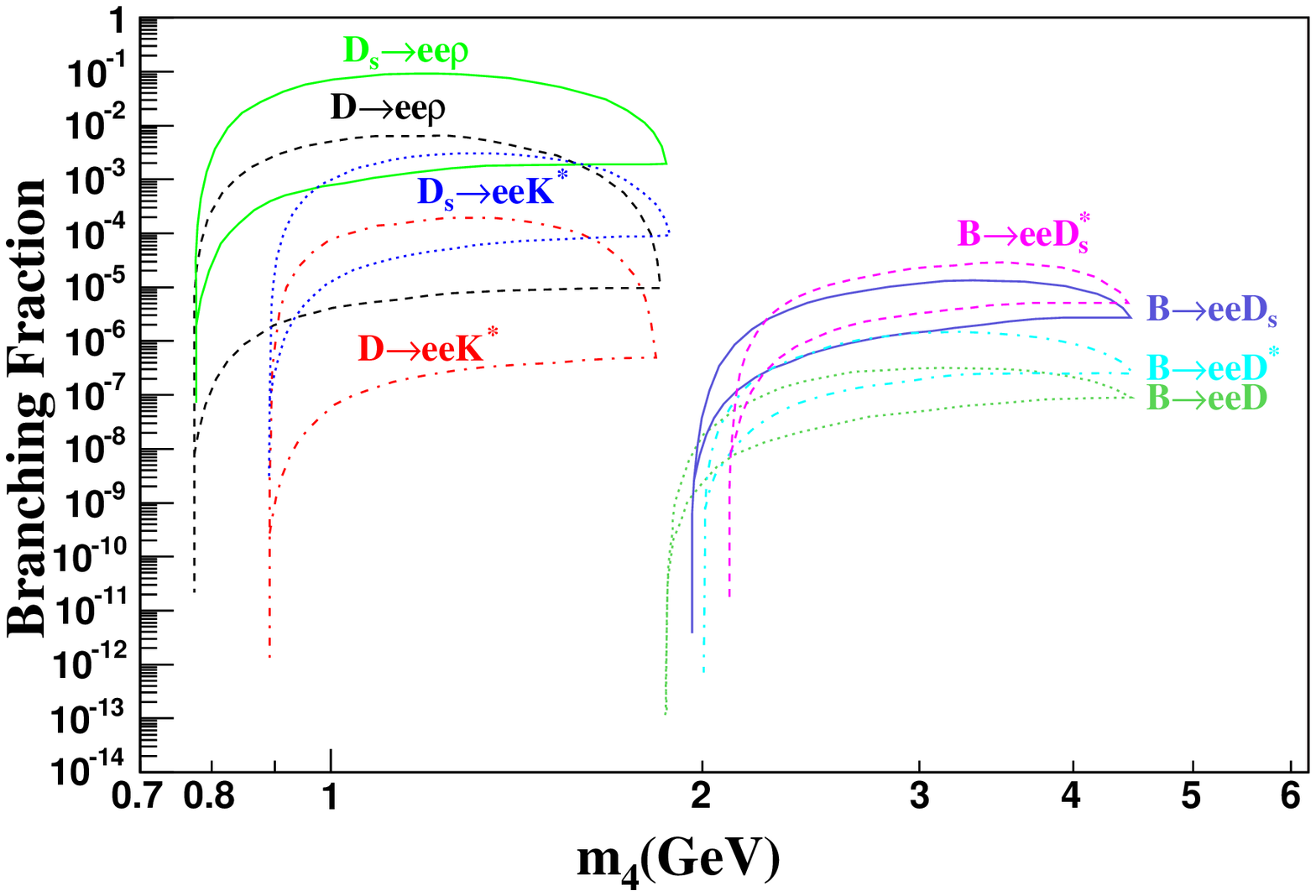}
\caption{\label{fig:ee} Theoretically excluded regions inside the
curve for the branching fraction of $M_1^+ \rightarrow e^+e^+M_2^-$
modes versus Majorana neutrino mass $m_4$. Regions below the curve
are theoretically allowed. The curves of $D_s \rightarrow
ee\rho$,~$D \rightarrow ee\rho$,~$D_s \rightarrow eeK^*$~and~$D
\rightarrow eeK^*$, denoted by solid line, break line, dotted line
and dash-dotted line, respectively.}
\end{figure}

\begin{figure}[tb]
 \centering
\includegraphics[width=5.0in]{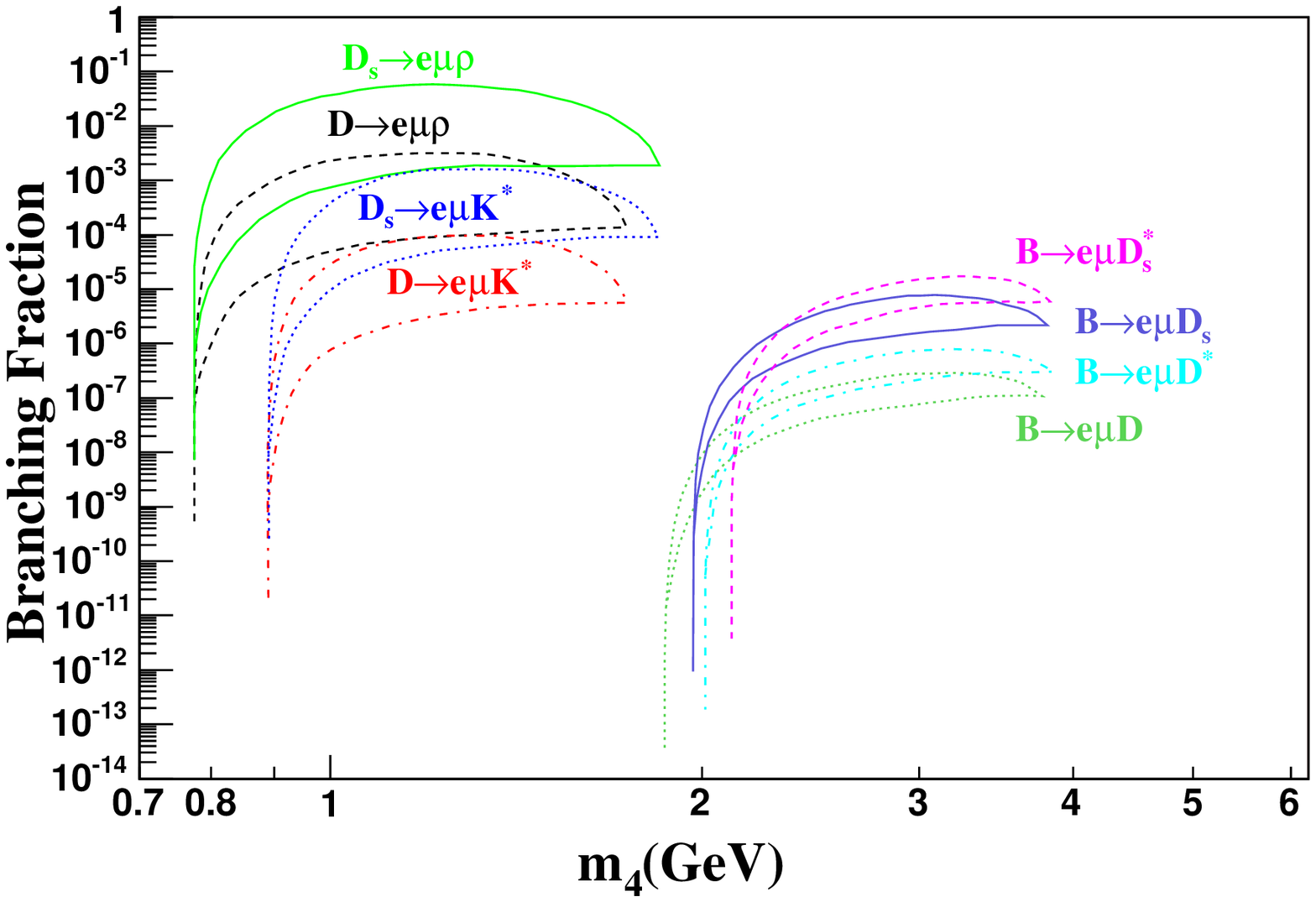}
\caption{\label{fig:eu} Theoretically excluded regions inside the
curve for the branching fraction of $M_1^+ \rightarrow
e^+\mu^+M_2^-$ modes versus Majorana neutrino mass $m_4$. Regions
below the curve are theoretically allowed. The curves of $D_s
\rightarrow e\mu\rho$,~$D \rightarrow e\mu\rho$,~$D_s \rightarrow
e\mu K^*$~and~$D \rightarrow e\mu K^*$, denoted by solid line, break
line, dotted line and dash-dotted line, respectively.}
\end{figure}

The key step to calculate decay widths and branching fractions of
the LV heavy meson decays is to determine the limits on the mixing
parameters $|V_{\ell_14}V_{\ell_24}|$ and neutrino mass $m_4$ in
Eq.~(\ref{ntr2}) and Eq.~(\ref{ntr3}). Generally speaking, one can
determine limits on the mixing parameters from the LV heavy meson
decay modes which have the current experimental limits on branching
fractions and determine the mass of neutrino by kinematics. However,
as mentioned in the introduction, since the LV decay modes which we
studied with vector meson and several pseudoscalar meson final
states are missing in directly experimental searches, so we cannot
yet get information of mixing parameters $|V_{\ell_14}V_{\ell_24}|$
from those decays. We thus propose the direct searches for those
modes in the existing and forth coming experiments such as in CLEO,
$B$-Factories, and the LHCb. On the other hand, there are direct
experimental results on the processes that may share common mixing
parameters with those under our consideration. These decay modes
given by Ref.~\cite{PDG04} and Ref.~\cite{CLEO1} have been
summarized and translated into the direct bounds in Ref.~\cite{Tao}.
We thus adopt them for consistency and carry out our analyses. We
first obtain the consistent limits on mixing parameters from some
decay modes which have the current experimental bounds on the
branching fractions for the heavy mesons with Ref.~\cite{Tao}. We
then translated the limits on mixing parameters to the relevant
decays modes of the heavy mesons $D$,\ $D_s$ and $B$ in the
Table~\uppercase\expandafter{\romannumeral1}. Depending on the
flavors of the final state leptons, the mixing parameters probed are
$|V_{e4}|^2$,\ $|V_{\mu4}|^2$ and $|V_{e4}V_{\mu4}|$, corresponding
to the decay modes:
\begin{equation}
M_1^+ \rightarrow e^+e^+M_2^-\,,\quad M_1^+ \rightarrow
e^+\mu^+M_2^-\quad {\rm and}\ \ M_1^+ \rightarrow \mu^+\mu^+M_2^-\,,
\label{eq:modes}
\end{equation}
respectively. We list the most stringent limits on $|V_{e4}|^2$,\
$|V_{\mu4}|^2$ and $|V_{e4}V_{\mu4}|$ for the 21 new decay modes in
Table~\uppercase\expandafter{\romannumeral1}. The ranges of heavy
neutrino mass $m_4$ in the
Table~\uppercase\expandafter{\romannumeral1} are determined by the
kinematics accessible.

When performing the calculations, the input parameters for the CKM
matrix elements and the decay constants of pseudoscalar and vector
mesons are chosen as follows \cite{PDG04,PDG08,CLEO2,MILC,Ebert}:
\begin{eqnarray}
&&|V_{ub}|=0.00359,\ |V_{cd}|=0.2256,\ |V_{cs}|=0.97334 ,\nonumber \\
&&f_{D^{\pm}}=0.2226\ {\rm GeV},\ f_{D_s^{\pm}}=0.266\ {\rm GeV},\
f_{B^{\pm}}=0.190\ {\rm GeV},\nonumber \\
&&f_{\rho^{\pm}}=0.220\ {\rm GeV},\ f_{K^{* \pm}}=0.217\ {\rm GeV},\
f_{D^*}=0.31\ {\rm GeV},\ f_{D_s^{*\pm }}=0.315\ {\rm GeV}.
\end{eqnarray}
We note that there may be some errors in determining the decay constants \cite{Error},
but they would not result in any qualitative difference for our predictions for
the SM-forbbiten modes.

\begin{figure}[tb]
 \centering
\includegraphics[width=5.0in]{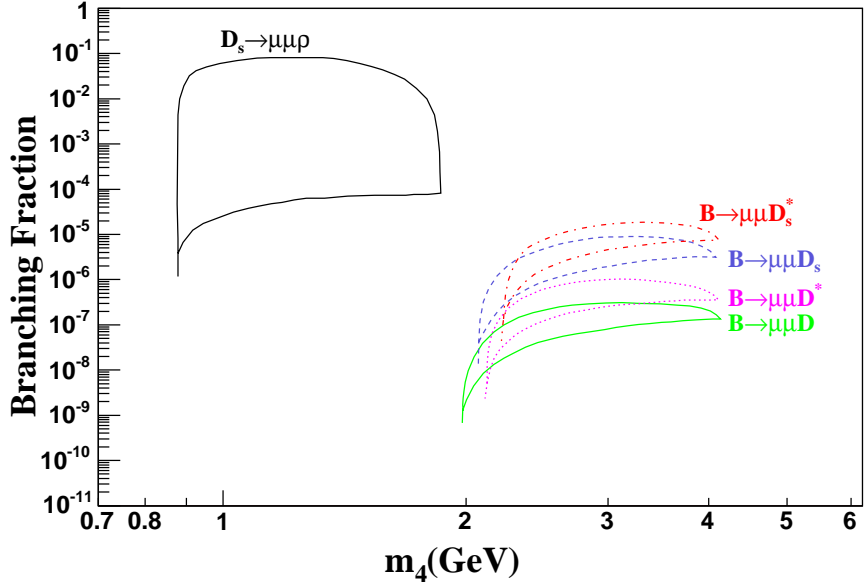}
\caption{\label{fig:uu} Theoretically excluded regions inside the
curve for the branching fraction of $M_1^+ \rightarrow
\mu^+\mu^+M_2^-$ modes versus Majorana neutrino mass $m_4$. Regions
below the curve are theoretically allowed.}
\end{figure}

With these parameters and the limits on mixing parameters and
corresponding mass ranges in the
Table~\uppercase\expandafter{\romannumeral1}, the decay widths and
branching fractions of the heavy mesons $D$,\ $D_s$ and $B$ are
calculated correspondingly. We perform a Monte Carlo sampling of the
branching fractions and the mass of the heavy neutrino, {\it i.e.},
we plot the excluded region of the branching fractions as a function
of $m_4$, as shown in Fig.~\ref{fig:ee} $\sim$ Fig.~\ref{fig:uu} for
the modes in Eq.~(\ref{eq:modes}). The regions inside the curve are
excluded by the direct experimental searches for the various LV
decay modes of heavy mesons as obtained in Ref.~\cite{Tao}. The
theoretical allowed regions are below the curve, {\it i.e.}, the
regions below the curve are the currently allowed branching
fractions for those LV heavy meson decay modes in the
Table~\uppercase\expandafter{\romannumeral1}.

From the figures, one can see that, if the heavy neutrino mass is
located in the range from $1$ GeV to $2$ GeV, even with the most
stringent constraints on mixing parameters, our theoretical
predictions of upper bound of the branching fractions can be large,
for example, $Br(D_s \rightarrow ee\rho)=1.8 \times 10^{-3}$,
$Br(D_s \rightarrow e\mu\rho)=2.0 \times 10^{-3}$ and $Br(D
\rightarrow e\mu\rho)=1.3 \times 10^{-4}$, since about $2.4 \times
10^6\ D^+D^-$ \cite{CLEO} events and $5.5 \times 10^5\
D_s^{*\pm}D_s^{\mp}$ \cite{CLEO3} events have been collected by CLEO
collaboration, these mentioned decay modes can be analyzed in the
current experiment, which will provide us a strong information of
$1$ GeV to $2$ GeV neutrino mass. But if the heavy neutrino mass is
heavier, around $2$ GeV to $4$ GeV, we cannot use the $D$ or $D_s$
decay modes to detect the LV processes because of the kinetic
limited, and the $B$ decay modes are favored, whereas our predicted
branching fractions of the LV $B$ decay channels are lower than $8.0
\times 10^{-6}$, which cannot be detected by current $B$-factories,
but in the forthcoming LHC experiments, around $10^{11} \sim
10^{12}$ $B$ meson events are expected \cite{Chen}, all the LV $B$
decays modes we studied can be effectively searched for by LHCb.

As a final remark, we would like to reiterate the advantage of our
treatment in searching for the resonant production and decay of a
Majorana neutrino in vector meson decay. Although kinematically
limited, the signal rate for the rare meson decay is substantially
enhanced due to the resonant nature. In contrast to the similar
decay channels as discussed in Ref.~\cite{Ali}, where the
intermediate Majorana neutrinos are far off mass-shell, the signal
would be much weaker. Other contributions such as the box diagrams
etc in Ref.~\cite{Ali} are all of similar nature and much smaller
than those considered here.

\section{Conclusions}

We extended the existing literature to consider the $\Delta L=2$
rare decays of heavy mesons $D,\ D_s,$ and $B$ to a vector or
pseudoscalar meson final state. Since there have not been any direct
experimental searches on these LV heavy meson decay modes, we
calculated their decay branching fractions and proposed to search
for them in the existing and forth coming experiments.

We first re-evaluated the limits on the mixing parameters $|V_{e4}|^2$,\
$|V_{\mu4}|^2$ and $|V_{e4}V_{\mu4}|$ from some decay modes which
have experimental limits on the branching fractions for
the heavy mesons, and obtained full agreement with those in
Ref.~\cite{Tao}. We then translated the limits on mixing parameters
and corresponding mass ranges to the relevant decays modes of $D,\ D_s,$ and $B$
of our current interests as summarized in
Table~\uppercase\expandafter{\romannumeral1}. Finally, we calculated
the decay widths and branching fractions for various LV decay modes
by the limits on the mixing parameters and the
heavy neutrino mass. We sampled the constraints on branching fractions
as a function of the heavy neutrino mass as shown in the figures.

Although the prevailing theoretical prejudice prefers Majorana
neutrinos, the unambiguous signature to prove the Majorana nature of
neutrinos is the experimental detection of a LV process. A detection
in one of the LV heavy meson decay modes studied in our analysis
would imply LV and hence the existence of a Majorana neutrino. At
present, about $2.4 \times 10^6\ D^+D^-$ \cite{CLEO} events and $5.5
\times 10^5\ D_s^{*\pm}D_s^{\mp}$ \cite{CLEO3} events have been
collected by CLEO collaboration. So these decay modes $D_s
\rightarrow ee\rho$, $D_s \rightarrow e\mu\rho$ and $D \rightarrow
e\mu\rho$ {\it et al.} which we studied in the
Table~\uppercase\expandafter{\romannumeral1} might show up in the
current experiments if the mass of the heavy neutrino is in the
range $1\ \rm GeV \lesssim m_4 \lesssim 2\ GeV$. But those $B$ decay
modes for the range of the heavy neutrino is $2\ \rm GeV \lesssim
m_4 \lesssim 4\ GeV$ cannot be detected presently due to small
branching fraction. Fortunately, in the forthcoming LHC experiments,
around $10^{11} \sim 10^{12}$ $B$ meson events are expected
\cite{Chen}, which will provide us with chances of discovering all
the LV $B$ decays modes we studied. Therefore, we may have the
opportunity to discover the LV process of heavy mesons $B$, $D$ and
$D_s$ via the distinctive channels of like-sign dilepton production
with no missing energy. Hadron colliders may serve as the discovery
machine for the mysterious Majorana neutrinos.

\begin{center}
\bf ACKNOWLEDGMENTS
\end{center}

We would like to thank Tao Han for his suggestions to carry out
this research, for providing the FORTRAN codes {\tt Hanlib} for the
calculations, and careful reading of the manuscript.
This work was supported in part by
the National Natural Science Foundation of China (NSFC) under Grant
No. 10875032 and in part by SRF for ROCS, SEM.

\end{document}